\documentclass[a4paper]{jpconf}

\usepackage{graphicx}              
\usepackage{amsmath}               
\usepackage{amsfonts}              
\usepackage{amsthm}   		         
\usepackage{color}	
\usepackage{hyperref}	
\usepackage{psfrag}
\usepackage[all, knot]{xy}
\usepackage{calc}

\newcommand{\C}{{\mathbb C}}

\newcommand{\R}{{\mathbb R}}

\newcommand{\cA}{{\mathcal A}}

\newcommand{\cH}{{\mathcal H}}
\newcommand{\cM}{{\mathcal M}}

\newcommand{\SU}{\mathrm{SU}}

\newcommand{\U}{\mathrm{U}}

\newcommand{\Spin}{\mathrm{Spin}}

\newcommand{\be}{\begin{equation}}
\newcommand{\ee}{\end{equation}}
\newcommand{\beq}{\begin{eqnarray}}
\newcommand{\eeq}{\end{eqnarray}}
\newcommand{\bes}{\begin{eqnarray}}
\newcommand{\ees}{\end{eqnarray}}
\newcommand{\bea}{\begin{eqnarray}}
\newcommand{\eea}{\end{eqnarray}}

\newcommand{\su}{{\mathfrak{su}}}

\renewcommand{\sl}{{\mathfrak{sl}}}
\newcommand{\la}{\langle}
\newcommand{\ra}{\rangle}

\newcommand{\f}{\frac}

\newcommand{\eps}{\epsilon}

\newcommand{\vV}{\vec{V}}

\newcommand{\id}{\mathbb{I}}

\def\arr{\rightarrow}

\newcommand{\matr}[2]{\left(\begin{array}{#1}#2\end{array}\right)}

\begin{document}

\title{Holomorphic Simplicity Constraints for 4d Riemannian Spinfoam Models}
\author{Ma\"it\'e Dupuis$^{1}$, Etera R. Livine$^2$}
\address{$^1$ Institute for Theoretical Physics III, University of Erlangen-N\"urnberg, Erlangen, Germany.}
\address{$^2$ Laboratoire de Physique, ENS Lyon, CNRS-UMR 5672, 46 All\'ee d'Italie, Lyon 69007, France.}
\ead{dupuis@theorie3.physik.uni-erlangen.de, etera.livine@ens-lyon.fr}

\begin{abstract}

Starting from the reformulation of the classical phase space of Loop Quantum Gravity in terms of spinor variables and spinor networks, we build coherent spin network states and show how to use them to write the spinfoam path integral for topological BF theory in terms of Gaussian integrals in the spinors. Finally, we use this framework to revisit the simplicity constraints reducing topological BF theory to 4d Riemannian gravity. These holomorphic simplicity constraints lead us to a new spinfoam model for quantum gravity whose amplitudes are defined as the evaluation of the coherent spin networks.

\end{abstract}


It was recently shown that the classical phase space for Loop Quantum Gravity on a fixed graph $\Gamma$ can be elegantly parameterized in terms of spinor networks \cite{twisted1, twisted2, un1, un2, spanishun1, spanishun2, spinorsEteraJohannes}. These spinor networks get quantized and lead to the spin network states of Loop Quantum Gravity.
This spinor formalism provides a clear interpretation of these states as discrete geometries. Here we use it to define coherent spin networks and build the spinfoam amplitudes for topological BF theory as Gaussian integrals over the spinor variables \cite{Fsimplicity2}. This allows to revisit the simplicity constraints reducing topological BF theory to 4d Riemannian gravity and to construct a spinfoam model for 4d Riemannian gravity based on new holomorphic simplicity constraints \cite{Fsimplicity1, Fsimplicity2}.

\section{Classical spinor networks and semi-classical spinnetworks} \label{SU2Spinors}

Spin network states for Loop Quantum Gravity can be seen as the quantization of classical spinor networks. These spinor networks are defined on a given graph $\Gamma$ and consist in spinor variables $z^v_e\in\C^2$ attached to the edges and vertices of $\Gamma$ and satisfying certain closure and matching constraints.

Let us first focus on a given vertex $v$ of $\Gamma$. For each edge $e$ attached to $v$, we define a spinor $|z_e\ra$ and its dual spinor, which are two-dimensional complex vectors living in the fundamental representation of $\SU(2)$:
\be
|z_e \ra = \matr{c}{z^0_e \\ z^1_e} \, \in \C^2,
\quad \la z_e |=\matr{cc}{\bar{z}_e^0& \bar{z}_e^1},
\quad \textrm{and} \quad
|z_e]=\varsigma |z_e \ra = \matr{c}{- \bar{z}^1_e \\ \bar{z}^0_e}\,.
\ee
The spinor space is naturally endowed with the canonical Poisson bracket
$
\{ z_e^a, \bar{z}_e^b \}=-i \delta_{ab}.
$
The spinors determine 3-vectors $\vec{V}(z_e)\in\R^3$, or $\vV_e$ for short, through their projection on Pauli matrices,
$
\vV(z_e)=\la z_e| \vec{\sigma} |z_e \ra.
$
This vector $\vV_e$ entirely determines the original spinor $z_e$ up to a global phase.
Its components form a $\su(2)$ algebra, $\{ V^i_e, V^j_e \}= 2 \epsilon^{ijk} V^k_e$,
and actually generates $\SU(2)$ transformation on the spinor, $|z_e\ra \,\arr\, g\,|z_e\ra$ with $g\in\SU(2)$.
We now impose the closure constraints $ \sum_e \vV_e=0$ at the vertex $v$, which generate a global $\SU(2)$ invariance on the spinors $z_e$. It reads in terms of the spinors as:
\be \label{closureSpinors}
\sum_e |z_e\ra \la z_e| = \f12 \sum_e \la z_e | z_e \ra \id.
\ee
$\SU(2)$-invariant observables are easily identified as scalar products between spinors as  $E_{ef}=\la z_e|z_f\ra$ and $F_{ef}=[z_e|z_f\ra$ \cite{un1,un2,spanishun2}. The modulus of these quadratic functionals give back the usual scalar products $\vV_e\cdot \vV_f$, or more precisely $|E_{ef}|^2=|\vV_e||\vV_f|+\vV_e\cdot \vV_f$ and $|F_{ef}|^2=|\vV_e||\vV_f|-\vV_e\cdot \vV_f$.

To form  a spinor network on the graph $\Gamma$, we glue together the structures defined at each vertex. We now have spinors $z^v_e$ around each vertex $v$, or equivalently two spinors $z^{s,t}_e$ for each edge attached $e$ to its source and target vertices, $s(e)$ and $t(e)$. These spinors satisfy the closure constraints at each vertex $v$ and new matching constraints on each edge \cite{twisted2,spanishun2}:
\be \label{gluingSpinors}
\la z_e^s | z_e^s \ra=\la z_e^t | z_e^t \ra
\quad \textrm{or equivalently}\quad
|\vV^s_e|=|\vV^t_e|\,.
\ee
These generate $U(1)$-transformations on the spinors, $z^{s,t}_e\arr \,e^{\pm i\theta}z^{s,t}_e$. The reduced phase space defined as the symplectic quotient $(\C^2)^{2E}//\SU(2)^V//\U(1)^E$ is exactly the phase space of Loop Quantum Gravity on the graph $\Gamma$. This is achieved explicitly by the reconstruction of the holonomy along the edges from the spinors \cite{twisted2,spanishun2,spinorsEteraJohannes}. The geometrical interpretation of these spinor networks is as a set of polyhedra (dual to each vertex) glued together by area-matching of their faces (but not shape-matching).

\section{Quantization and Coherent Spin Networks}

The quantization is straightforward and the spinor components are raised to annihilation and creation operators of harmonic oscillators:
\be
| z_e \ra= \matr{c}{z^0_e \\ z^1_e}
\, \rightarrow \,
\matr{c}{a_e \\ b_e},
\qquad
\la z_e|=\matr{cc}{\bar{z}_e^0& \bar{z}_e^1}
\,\rightarrow \,
\matr{cc}{a^\dagger_e& b^\dagger_e},
\ee
with $[ a_e, a_e^\dagger]=[ b_e, b_e^\dagger]=1$ and $[a_e, b_e]=0$ for all $e$. This leads to Schwinger representation of $\SU(2)$ in terms of a couple of harmonic oscillators. The components of the vector $\vV_e$ become the $\su(2)$ generators and the spin $j_e$ of the corresponding irreducible $\SU(2)$-representation is given by fixing the total energy $E_e=|\widehat{\vV(z_e)}|=(a_e^\dagger a_e +b_e^\dagger b_e )=2j_e$.
At the quantum level, the closure constraints impose $\SU(2)$-invariance at each vertex $v$, while the matching constraints impose that the spins at the source and target vertices of each edge are equal. At the end, we recover spin network states, with $\SU(2)$-representation $j_e$ on each edge $e$ and intertwiners at each vertex $v$, that is an $\SU(2)$-invariant state in the tensor product of the $\SU(2)$-representations $j_e$ living on the edges $e$ of $v$.

Next we introduce coherent spin networks peaked on each point of the spinor phase space. Starting with intertwiners, we define coherent intertwiners\footnote{There is a clear relation between these coherent states and the LS coherent intertwiners used in the definition of the EPRL-FK spinfoam models \cite{FK, LS}. The interested reader can find details in \cite{un2,Fsimplicity2}.} as the group averaging of the standard  coherent states for the harmonic oscillators labeled by the spinors $\{z_e\}$ \cite{Fsimplicity1, Fsimplicity2}:
\be \label{SU2coherent}
||\{z_e \}\ra \equiv \int dg \, g \triangleright  e^{ \sum_e z_e^0 a^\dagger_e + z_e^1 b_e^\dagger} | 0 \ra,
\ee
where the vacuum state $|0 \ra$ is the vacuum of the harmonic oscillators. These coherent states transform nicely under $\U(N)$ (see \cite{un2,Fsimplicity2} for details) and provide a decomposition of the identity on the Hilbert space $\cH_N$ of  $N$-valent intertwiners:
\be \label{identity}
\id_{\cH_N}= \f{1}{\pi^{2N}} \int [d^4z_e]^N e^{-\la z_e| z_e \ra} ||\{z_e\}\ra \la \{ z_e \}Ê||.
\ee
Then we glue these coherent intertwiners to form a full coherent spin network on the graph $\Gamma$.  We have intertwiners $|| \{ z_e^v \}_{e\ni v} \ra$ for each vertex $v \in \Gamma$ where the set of spinors $\{z_e^v\}$ satisfy the closure and  matching constraints. The coherent quantum state on $\Gamma$ is defined as the tensor product of these intertwiner states, $\psi_{\{z_e^v\}}=\bigotimes_v ||\{z^v_e\}_{e\ni v}\ra$. $\psi_{\{z_e^v\}}$ is truly labeled by points in the reduced phase space and its evaluation on the group elements $\{g_e\} \in \SU(2)^E$  reads:
\beq \label{coherentspinnet}
\psi_{\{z_e^v\}}(g_e)
\,=\,
\int [dh_v]\,
e^{\sum_e [ z_e^{s(e)}|h_{s(e)}^{-1}\,g_e\,h_{t(e)}|z_e^{t(e)}\ra}\,.
\eeq
This expression is explicitly  $\SU(2)$-invariant at every vertex $v$ and holomorphic in the spinors $z_e^v$'s. These coherent spin networks are semi-classical states peaked on the discrete geometry defined by their classical spinor labels \cite{Fsimplicity1,Fsimplicity2}.
%

\section{Spinfoam amplitudes for BF theory in terms of spinors}

At this stage we have all the tools necessary to re-write the spinfoam path integral for 4d BF theory with $\SU(2)$ as gauge group in terms of spinors and holomorphic coherent intertwiners. Generic spinfoams are defined on arbitrary 2-complexes, describing the time evolution of graphs. Representations (resp. intertwiners) are put on faces (resp. edges) on the 2-complexes and the spinfoam amplitude is defined as the integral over all possible data of the product of local amplitudes associated to each vertex of the 2-complex. The neighborhood of a vertex defined by the incoming edges and faces dressed with intertwiners and representations is described by a boundary graph and spin network states and the local vertex amplitude is defined as the evaluation of this boundary spin network (at the identity).
Here, for simplicity's sake, we will only consider 4d triangulations made of 4-simplices glued together along tetrahedra, with the 2-complex defined as the dual 2-skeleton and the spinfoam vertices as dual to the 4-simplices.
Considering on a single 4-simplex $\sigma$ and its dual boundary graph, we attach coherent intertwiners to each dual vertex (i.e tetrahedron) and we define the 4-simplex amplitude as the evaluation of the corresponding coherent spin network i.e $\psi_{\{z_\Delta^\tau\}}(g_\Delta) \equiv \psi_{\{z_e^v\}}(g_e)$ defined by \eqref{coherentspinnet} evaluated at $g_e=\id$.
$\tau$ label the five tetrahedra of the 4-simplex and are dual to the vertices $v$ of the  boundary graph, while $\Delta$ denote the ten triangles dual to the edges $e$ of the boundary graph.
Then the 4-simplices are glued together using the resolution of the identity on the space of intertwiners. There is nevertheless an ambiguity in the integration measure for the spinors and we fix it by requiring that the spinfoam amplitude for the $\SU(2)$ BF theory be topological, which gives a Gaussian integral over the spinor variables \cite{Fsimplicity2}:
$$
Z[\cM]
=
\int \prod_{\tau,\Delta\in\tau}
\f{e^{-\la z^\tau_\Delta|z^\tau_\Delta\ra}d^4z_\Delta^\tau}{\pi^2}\,
\prod_\Delta \mu(z_\Delta^{\tau(\Delta)})
\prod_\sigma\int \prod_{\tau\in\sigma}dh_\tau^\sigma\,
e^{\sum_{\Delta\in\sigma}
[\varsigma^{\eps^{\sigma}_{s(\Delta)}}\, z_\Delta^{s(\Delta)}
|(h^\sigma_{s(\Delta)})^{-1}h^\sigma_{t(\Delta)}|
\varsigma^{\eps^{\sigma}_{t(\Delta)}}\,z_\Delta^{t(\Delta)}\ra}
$$
where the measure factor $\mu(z)=(\la z|z\ra-1)$ is inserted for a single tetrahedron $\tau(\Delta)$ for each triangle $\Delta$ (the origin of the dual plaquette) and the signs $\eps^\sigma_\tau$ register the orientations of the tetrahedra with respect to each 4-simplex.
Using the decomposition of the $\delta$-distribution of $\SU(2)$ as an integral over spinors,
\be
\delta(g_1..g_n)=\int\f{e^{-\la z|z\ra}d^4z}{\pi^2}\mu(z)\, e^{\la z|g_1..g_n|z\ra}
=\int \f{\prod_i^n e^{-\la z_i|z_i\ra}d^4z_i}{\pi^{2n}}\mu(z_1)\,\, e^{\la z_1|g_1|z_2\ra}..e^{\la z_n|g_n|z_1\ra},
\ee
we show that the definition above reproduces exactly the discretized path integral for BF theory:
\be
Z[\cM]
\,=\,
\int \prod_{\sigma,\tau\in\sigma}dh_\tau^\sigma\,
\prod_\Delta \delta(\overrightarrow{\prod}_{\sigma\ni \Delta} (h^\sigma_{s(\Delta)})^{-1}h^\sigma_{t(\Delta)})\,.
\ee

\section{Holomorphic simplicity constraints for the Euclidean case} \label{SF}

We now go from the topological BF theory for $\Spin(4)\sim \SU_L(2)\times \SU_R(2)$ to 4d Riemannian gravity within the spinorial framework. The simplicity constraints are imposed at level of each intertwiner (or tetrahedron or edges of the dual 2-complex). Working at a vertex $v$ with now two sets of  spinors $\{z_e^{v,L}\}$ and $\{z_e^{v,R}\}$ independently satisfying the closure constraint, we introduce new  simplicity constraints (where we have dropped the index $v$):
\be \label{simplicity}
\forall e,f\,\quad F^L_{ef}=\rho^2 F^R_{ef}
\qquad
\textrm{i.e.}
\quad
\forall e,f\,\quad [z_e^L|z_f^L\ra=\rho^2 [z_e^R|z_f^R\ra
\ee
where $\rho$ is  a fixed parameter related to the Immirzi parameter $\gamma$.
At the classical level, these holomorphic simplicity constraints with the closure constraints imply the usual quadratic simplicity constraints (by taking their modulus square). They are also equivalent to the linear simplicity constraints proposed in \cite{EPR}, i.e the existence of a 4-vector normal to all the bivectors $B_e+ \gamma \star B_e$ where $B_e$ is defined by its self-dual and anti-self dual parts $(\vV(z_e^L),\vV(z_e^R))$ \cite{Fsimplicity2}.
At the quantum level, the coherent intertwiner $||\{z_e\}\ra$ diagonalizes the $\hat{F}$ annihilation operators, $\hat{F}_{ij}\,||\{z_e\}\ra=[z_i|z_j\ra\,||\{z_e\}\ra$. Thus we solve exactly the holomorphic simplicity constraints by the coherent intertwiners $|\{z_e\}\ra_\rho\,\equiv\,   ||\{\rho z_e\}\ra_L\otimes||\{z_e\}\ra_R$
with $z_e^L=\rho z_e=\rho z_e^R$ \cite{Fsimplicity1, Fsimplicity2}.
%
%
Then we define a new spinfoam model solving exactly the holomorphic simplicity constraints by its vertex amplitude given by the evaluation of the coherent spin network on the boundary 4-simplex graph obtained by gluing these coherent simple intertwiners $|\{z_e\}\ra_\rho$:
\be
{}_\rho\cA_\sigma(z_\Delta^\tau)=
\psi_{\{\rho z_\Delta^\tau\}}(\id)\psi_{\{z_\Delta^\tau\}}(\id)=
\int [dh_\tau]^5\,e^{\sum_{\Delta\in\sigma}
\rho^2[z_\Delta^{s(\Delta)}|h^L_{s(\Delta)}{}^{-1}h^L_{t(\Delta)}|z_\Delta^{t(\Delta)}\ra
[z_\Delta^{s(\Delta)}|h^R_{s(\Delta)}{}^{-1}h^R_{t(\Delta)}|z_\Delta^{t(\Delta)}\ra}
\,.
\ee
The full spinfoam amplitude is obtained by gluing these vertex amplitudes and integrating over the spinors with a Gaussian measure. The main difference with the EPRL-FK spinfoam model is that we do have quantum states solving the holomorphic simplicity constraint operators. The ``drawback" is that the diagonal simplicity constraints are not strongly imposed and we do not work with simple representations satisfying $j_e^L=j_e^R$, but with Gaussian wave-packets peaked on this relation.

Finally, the reformulation of spinfoam amplitudes in terms of spinors presents the advantage of a clear link with coherent spin network states and an expression directly as the path integral of a discrete action in terms of spinors and holonomies. We hope it could be an interesting starting point for the study of the renormalization and coarse-graining of the spinfoam amplitudes.



\end{document}